# Spectrophotometry of globular clusters in NGC 5128(Cen A): determination of their metallicities

P. Jablonka[1,2**], E. Bica[3], D. Pelat[2], and D. Alloin[2]

[1] European Southern Observatory, Karl Schwarzschild Strasse 2. 85748, Germany
[2] URA 173 CNRS-DAEC, Observatoire de Paris, 92195 Meudon Principal Cedex France
[3] Departamento de Astronomia, IF-UFRGS, CP 15051, CEP 91501-970, Porto Alegre, RS, Brazil

**Abstract.** We have carried out spectrophotometry in the range 3600–9700Å of five globular clusters in NGC 5128 for which previous photometric studies suggested a high metal content. We compare the equivalent widths of a set of metallic features of the NGC 5128 clusters with those of well-studied Galactic and M31 globular clusters. This enables us to derive a reliable ranking and estimate of the metallicity for the NGC 5128 clusters. The reddening is obtained from the comparison of the continuum distribution of the observed clusters with those of reference clusters. The present spectroscopic study clarifies the question raised in the photometric studies about the NGC 5128 clusters metallicities. We conclude that, in the present NGC 5128 sample, no object has actually a metallicity higher than solar, unlike what happens in M31.

## 1. Introduction

Star clusters are valuable tools for the analysis of stellar populations in galaxies. They are basically formed by stars of a same age and metallicity, which allows one to infer on the formation conditions and subsequent evolution of the parent galaxy. In order to understand such processes, it is important to compare properties, such as metallicity, kinematics and spatial distribution, of globular cluster systems in different galaxies, and to see how they correlate with the parent galaxy characteristics (see e.g. Harris 1991 for a review). Since the average metallicity of a globular cluster system appears to increase with the parent galaxy mass (van den Bergh 1975; Huchra, Brodie & Kent 1991), the closest giant early-type galaxy NGC 5128 (Centaurus A) is of great interest to search for extremely metallic globular clusters. The integrated B−V (Hesser et al. 1984) and V−K colours (Frogel 1984) suggest not only that the mean metallicity of the globular cluster system of NGC 5128 is higher than that in our Galaxy or in M31, but also that some of these clusters might have metallicities well above the solar value. More recently, by means of integrated photometry in the Washington system, Harris et al. (1992) (hereafter HGHH92) confirm the high mean metallicity of the globular cluster system; however, they find evidence that the metallicities of the reddest clusters are not as extreme as estimated by Frogel (1984). The issue requires spectroscopic observations to be settled.

In the present study, we analyze spectroscopically a selection of globular clusters in NGC 5128, based on indications of their very high metallicity, as deduced from the photometric studies described above, with the aim of clarifying the extreme metallicity issue and to compare such clusters with globular clusters in the Local Group. In order to disentangle age, metallicity and reddening effects, it is essential to observe over a wide spectral range for a good continuum baseline and an extensive series of fundamental spectral features. It is also important to use calibration indices sensitive to the total range of metallicity observable in galaxies. A database of star cluster spectra over the range 3500 Å–10 000 Å in our Galaxy, M31 and the Magellanic Clouds are available for the present analysis (Bica & Alloin, 1986a, b, 1987; Jablonka, Alloin & Bica 1992). Our database includes the central clusters G177 and G158 in M31, which present metallic absorption features similar to those observed in the nuclei of giant ellipticals (Jablonka, Alloin & Bica 1992 ; Bica et al. 1992), thus well-suited for comparisons with extreme metallicity clusters in NGC 5128, if any.

---

discuss the appearance of the spectra as compared to key clusters in the Galaxy and in M31, and provide the measurements of equivalent widths and of continuum points. In Sect. 4 reddening and metallicity values are estimated and the results are discussed. Our concluding remarks are given in Sect. 5.

## 2. Observations

We have selected five clusters, namely #3, #11, #12, #23, and #26 (numbers following Hesser, Harris & Harris 1986, hereafter HHH86) in NGC 5128. Frogel (1984) estimated their metallicies to be in the range $+0.23 \leq$ [Fe/H] $\leq +0.91$. However, for three of the metal-rich clusters in our sample (#11, #12 and #23), HGHH92 derived considerably lower metallicities ([Fe/H]=$-0.2$, $-0.1$ and $-0.1$ respectively).

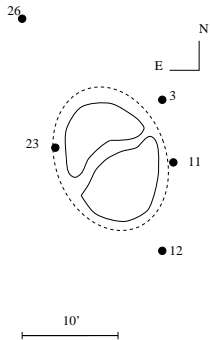

**Fig. 1.** Spatial distribution of globular clusters in NGC 5128

The clusters were observed during a run of three nights in April 1993 at the ESO/La Silla 3.60 m telescope, equipped with EFOSC in the long-slit spectroscopic mode. The logbook of the observations is summarized in Table 1. The slit was set tangentially to the galaxy's isophotes in order to minimize the background variations. The slit width was 1.5″, basically gathering most of the integrated light of each cluster. A Tek512 ESO #26 CCD with 512×512 pixels of size 27$\mu$m was employed. We have used gratings B300 (6.3 Å/pixel) and R300 (7.5 Å/pixel), spanning respectively the ranges 3600–7000Å and 5600–9900Å. The position angles (PA) and the average air masses (AM) in each spectral range are indicated in Table 1. Three exposures of 30 minutes each were taken for each spectral range. The reductions were carried out in a standard way at ESO/Garching with the IRAF 2.10 long-slit package. We corrected the cluster spectra for redshift by taking the velocity values given in HHH86. Uncertainties in the redshift correction are small as compared to the spectral windows used in the equivalent width measurements. We subsequently linked the observed visible and near-infrared spectral ranges. A global signal-to-noise ratio of ~30 has been reached.

**Table 1.** Logbook of the observations

| Cluster | PA | AM (Blue & Red) |
|---|---|---|
| 3 | -43° | 1.57 & 1.14 |
| 11 | 90° | 1.03 & 1.20 |
| 12 | 10° | 1.13 & 1.03 |
| 23 | 80° | 1.13 & 1.06 |
| 26 | 0° | 1.20 & 1.04 |

We gather in Table 2 information on the clusters: in column (1), the cluster number; in column (2), the angular distance R in arc minutes from the center of NGC 5128 (HHH86); in columns (3) and (4), the observed photographic (B−V) colours from Hesser et al. 1984 and HGHH92 respectively; in columns (5), (6) and (7), the Washington system colours from HGHH92; in column (8) and (9), the infrared colours from Frogel (1984); and finally, in column (10) and (11), the metallicities from Frogel (1984) and HGHH92 respectively. We show in Fig. 1 the spatial distribution of our cluster sample, based on the plate in HHH86. The objects are located outside the galaxy's main body, which favors their observation.

## 3. Results

### 3.1. Spectral Appearance

We show, in Fig. 2a and Fig. 3a, the spectra of the observed clusters in NGC 5128; they have *not* been corrected for foreground reddening. In Fig. 2b and Fig. 3b, we show, for comparison, some reference spectra of globular clusters in our Galaxy and in M31. G1 and G2 are templates from Bica (1988), where the blue range is replaced by the CCD spectra taken by Bica, Alloin & Schmidt (1994). G1 is an average of the nearly-solar metallicity clusters NGC 6440, NGC 6528 and NGC 6553; G2 averages a series of globular clusters around $[Z/Z_{\odot}] \sim -0.5$. The important reference cluster 47 Tuc is located at the low-metallicity end of the group of objects averaged to form the template spectrum G2. The M31 star cluster G177 appears to be a genuine super-metal-rich globular cluster near the nucleus of M31 (Jablonka, Alloin & Bica 1992 ; Bica et al. 1992). The comparison spectra are corrected for foreground reddening using a normal Galactic law (Seaton 1979) and E(B−V) values as described in our references.

A simple inspection of Figures 2 and 3 suggests that the metal-rich clusters in NGC 5128 are more metallic than the template G2 and less than the super-metal-rich cluster G177 in M31.

### 3.2. Measurements

We measured the equivalent widths of spectral absorption features and the continuum distribution following window limits and continuum tracings from Bica & Alloin

**Table 2.** Information about the NGC 5128 clusters, from the literature

| (1) | (2) | (3) | (4) | (5) | (6) | (7) | (8) | (9) | (10) | (11) |
|---|---|---|---|---|---|---|---|---|---|---|
| Cluster | R(') | B−V | B−V | C−M | M−$T_1$ | $T_1$−$T_2$ | J−H | H−K | [Fe/H]$_{V-K}$ | [Fe/H]$_{C-T_1}$ |
| 3 | 7.2 | 1.02 | − | − | − | − | 0.96 | 0.19 | +0.35 | − |
| 11 | 5.8 | 1.12 | 1.14 | 1.158 | 0.905 | 0.705 | 0.96 | 0.24 | +0.51 | 0.0 |
| " | " | " | " | 1.118 | 0.827 | 0.749 | " | " | " | −0.4 |
| 12 | 10.1 | 1.06 | 1.08 | 1.155 | 0.870 | 0.723 | 0.96 | 0.22 | +0.81 | −0.1 |
| 23 | 4.9 | 1.12 | 1.18 | 1.152 | 0.870 | 0.731 | 0.89 | 0.23 | +0.23 | −0.1 |
| 26 | 14.9 | 1.16 | − | − | − | − | 0.99 | 0.27 | +0.91 | − |

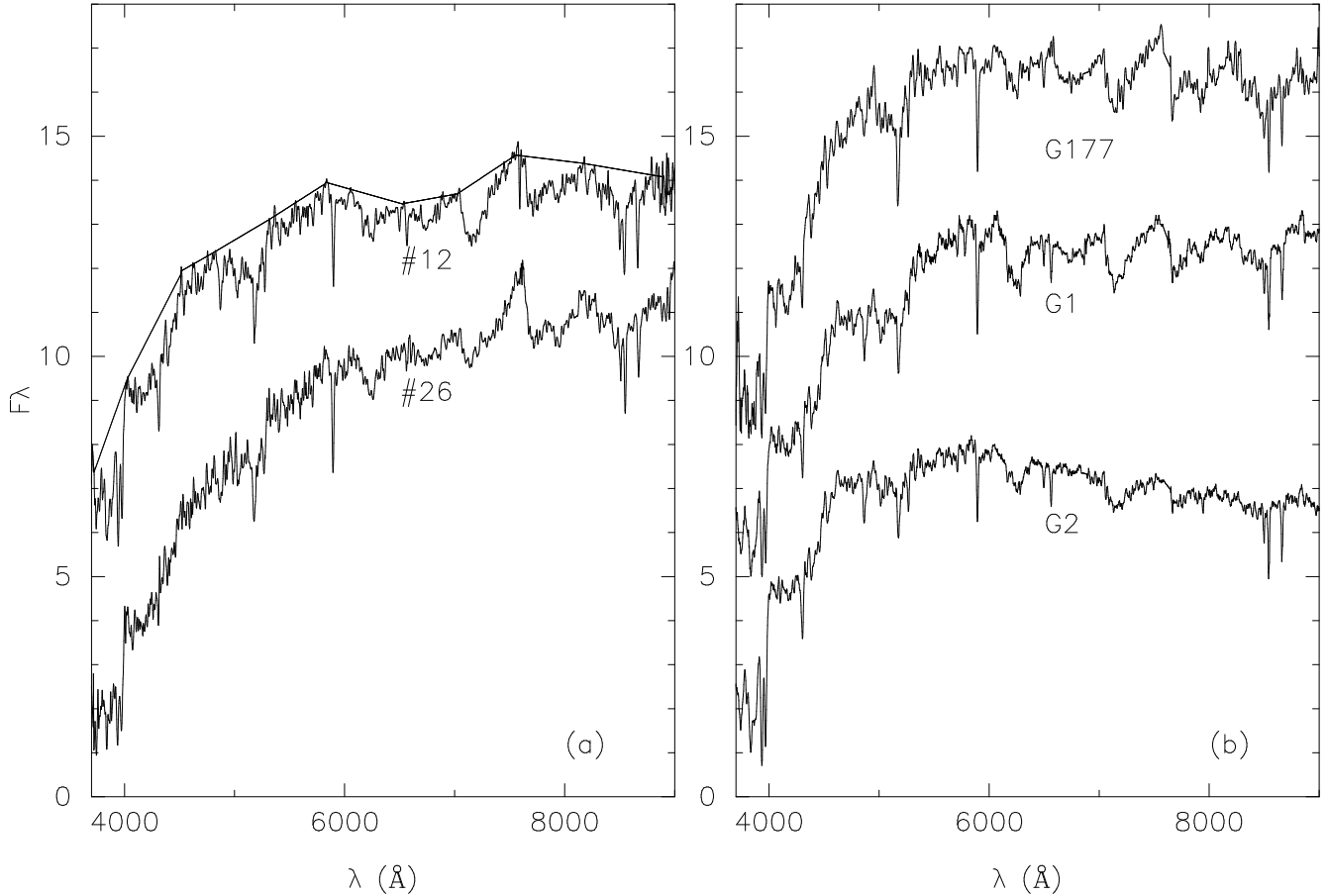

**Fig. 2.** Comparison of the most strong-lined spectra in our sample of NGC 5128 clusters in (*a*) with reference clusters in our Galaxy and M31 in (*b*). As an example, we display the chosen continuum of the NGC 5128 #12 cluster.

(1986a,b) and Bica & Alloin (1987). A continuum tracing is illustrated on one of the NGC 5128 cluster spectra in Fig. 2a. We recall, in Table 3, the main absorbers and the limits of the spectral windows, and show the results of the measurements for the NGC 5128 star clusters, as well as for the comparison objects. We also include in Table 3 the template G3 (Bica 1988), which is an average of globular clusters around [Z/Z$_\odot$]≈−1.0. Table 4 provides the measurements of a set of continuum points, where those of the NGC 5128 clusters have not been corrected for any reddening, while those of the comparison objects have been corrected.

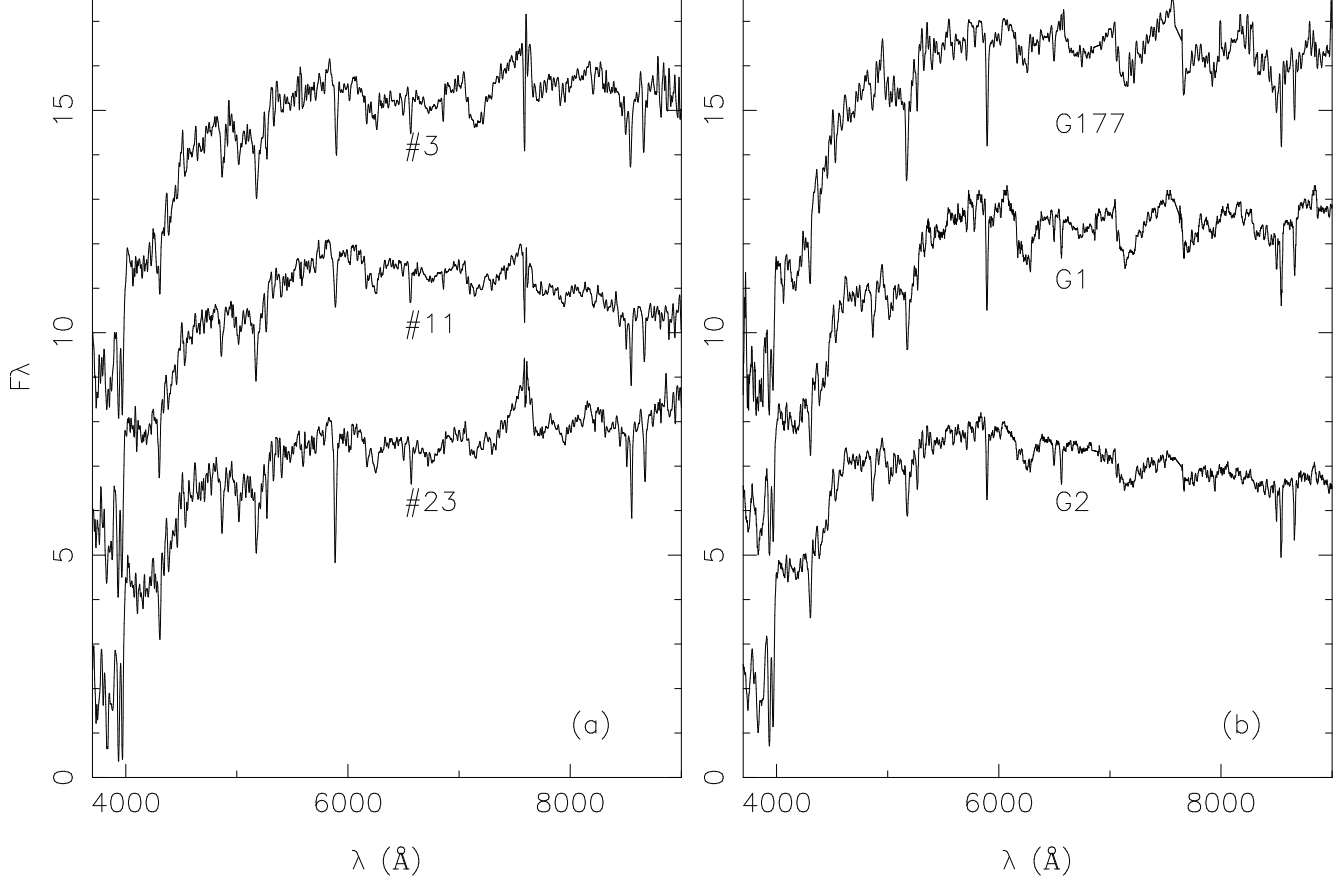

**Fig. 3.** Comparison of the weak-lined spectra in our sample of NGC 5128 clusters in (a) with reference clusters in our Galaxy and M31 in (b).

## 4. Discussion

### 4.1. Metallicity

In order to evaluate the ranking and absolute metallicity values for the five globular clusters in NGC 5128, we simply add all the equivalent widths, W, of the metallic features in Table 3. Thus, we use 23 spectral windows out of the 27 which have been measured, eliminating those which are dominated by Balmer lines. The absorbers in the metallic windows include CNO and heavier elements such as Mg, Ca and Fe, so that the resulting sum of equivalent widths, $\Sigma W$, should be representative of the total abundance of metals, Z. The resulting values (Table 5) are used to rank the NGC 5128 clusters in metal content among themselves and with respect to the reference clusters. Cluster #26 is the most metallic in our sample, followed by #12, in agreement with the ranking of Frogel (1984) for these clusters. However, #26 hardly reaches the level of the metal-rich Galactic cluster template, G1, in terms of $\Sigma W$, and G177 in M31 is by far more strong-lined than #26. Cluster #11, the least strong-lined cluster in our sample, has $\Sigma W$ slightly larger than that of the template G2 and consequently it is more metallic than 47 Tuc in the Galaxy.

**Table 5.** Sum of W for 23 metallic windows

| Cluster | $\Sigma W$(Å) | Template | $\Sigma W$(Å) |
|---|---|---|---|
| 3 | 209.7 | G3 | 95.7 |
| 11 | 187.2 | G2 | 179.8 |
| 12 | 219.5 | G1 | 244.2 |
| 23 | 205.6 | G177 | 277.0 |
| 26 | 234.0 | | |

An absolute metallicity scale for the metal-rich globular clusters is not yet settled. In consequence, we estimate metallicity values of the NGC 5128 clusters by, first assuming the templates G1 and G2 at $[Z/Z_\odot]=0.0$ and $-0.5$ respectively, and then interpolating linearly the values in Table 5. The assumption of a solar metal abundance for G1 is consistent with the recent results of Barbuy et al. (1992). They derive an abundance of heavy met-als [M/H]=−0.2 and [N/Fe]=+0.4 for a giant star in NGC 6553, by means of high dispersion spectrophotometry. Our final estimates are shown in Table 6. As the average errors in W values amount to ≈5%, this implies a formal error in metallicity of $\epsilon$([Z/Z$_\odot$])=±0.08 (thus excluding uncertainties attached to the absolute metallicity scale itself). In the assumed scale, the super metal-rich cluster G177 has [Z/Z$_\odot$]=+0.25. The most metal-rich cluster in our sample of NGC 5128 clusters, #26, is slightly subsolar in such a scale.

**Table 3.** Equivalent Widths

| Window | Absorber | Limits (Å) | NGC 5128 Globular Clusters | | | | | Comparison Objects | | | |
|---|---|---|---|---|---|---|---|---|---|---|---|
| | | | 3 | 11 | 12 | 23 | 26 | G3 | G2 | G1 | G177 |
| 2 | CN, H9 | 3814-3862 | 20.5 | 19.8 | 22.7 | 18.9 | 22.4 | 8.1 | 18.8 | 22.3 | 26.5 |
| 3 | CN, H8 | 3862-3908 | 13.8 | 13.3 | 16.6 | 14.8 | 17.4 | 6.5 | 14.1 | 16.3 | 20.2 |
| 4 | Ca II K | 3908-3952 | 18.5 | 17.9 | 17.7 | 16.9 | 20.6 | 10.0 | 16.3 | 17.6 | 21.0 |
| 5 | Ca II H,H$\epsilon$ | 3952-3988 | 14.0 | 13.2 | 13.4 | 14.0 | 18.1 | 9.3 | 12.5 | 13.4 | 16.3 |
| 9 | H$\delta$ | 4082-4124 | 4.9 | 3.3 | 4.7 | 5.3 | 5.0 | 2.9 | 2.7 | 4.2 | 7.0 |
| 10 | Fe I | 4124-4150 | 4.3 | 3.3 | 3.5 | 3.9 | 4.3 | 1.0 | 2.2 | 4.0 | 6.9 |
| 11 | CN | 4150-4214 | 10.0 | 9.7 | 10.9 | 10.4 | 13.3 | 2.9 | 7.9 | 12.5 | 19.0 |
| 12 | Ca I | 4214-4244 | 4.8 | 4.2 | 4.4 | 4.1 | 5.1 | 1.7 | 3.7 | 5.1 | 7.0 |
| 13 | Fe I | 4244-4284 | 7.4 | 5.7 | 6.6 | 6.3 | 7.6 | 2.9 | 5.1 | 6.8 | 9.7 |
| 14 | CH G | 4284-4318 | 8.8 | 8.8 | 9.7 | 9.3 | 7.1 | 5.7 | 8.2 | 9.7 | 10.8 |
| 15 | H$\gamma$ | 4318-4364 | 6.1 | 5.4 | 6.4 | 6.0 | 6.2 | 4.7 | 5.6 | 6.2 | 5.6 |
| 16 | Fe I | 4364-4420 | 8.3 | 7.1 | 8.2 | 7.0 | 7.6 | 4.3 | 7.9 | 10.4 | 9.4 |
| 27 | H$\beta$ | 4846-4884 | 4.2 | 3.5 | 4.2 | 4.1 | 4.2 | 3.2 | 3.9 | 4.7 | 3.9 |
| 31 | Fe I | 4998-5064 | 6.2 | 5.5 | 7.1 | 6.2 | 5.2 | 2.6 | 5.2 | 8.0 | 6.8 |
| 32 | Fe I, $C_2$ | 5064-5130 | 5.2 | 4.2 | 6.4 | 4.7 | 4.8 | 1.9 | 4.3 | 6.7 | 8.3 |
| 33 | MgH, $C_2$ | 5130-5156 | 3.1 | 2.6 | 3.3 | 2.7 | 2.8 | 1.2 | 2.3 | 3.2 | 3.8 |
| 34 | Mg I, MgH | 5156-5196 | 7.7 | 7.0 | 8.8 | 7.3 | 8.3 | 3.0 | 5.9 | 8.4 | 9.8 |
| 35 | MgH | 5196-5244 | 4.9 | 4.0 | 6.2 | 5.1 | 5.5 | 1.5 | 3.8 | 6.0 | 5.0 |
| 36 | Fe I | 5244-5314 | 4.7 | 3.1 | 5.9 | 4.5 | 4.9 | 1.6 | 4.5 | 5.0 | 4.3 |
| 48 | Na I | 5880-5914 | 4.4 | 3.1 | 4.6 | 6.1 | 5.3 | 1.7 | 3.6 | 5.2 | 5.8 |
| 54/5/6/7 | TiO | 6156-6386 | 13.4 | 12.5 | 14.8 | 13.4 | 17.2 | 8.1 | 12.9 | 21.6 | 15.9 |
| 60 | H$\alpha$ | 6540-6586 | 2.2 | 2.2 | 2.5 | 2.0 | 1.3 | 2.7 | 1.8 | 1.9 | -0.6 |
| 66/7/8 | TiO | 7050-7464 | 29.7 | 21.5 | 28.0 | 24.3 | 29.1 | 6.2 | 19.0 | 29.7 | 38.7 |
| 75/76 | TiO,Ti | 8234-8476 | 7.5 | 7.3 | 7.7 | 10.8 | 11.9 | 7.0 | 9.9 | 15.8 | 14.1 |
| 77 | Ca II,TiO | 8476-8520 | 3.3 | 3.5 | 3.9 | 3.8 | 4.7 | 2.1 | 3.1 | 4.9 | 6.0 |
| 78 | Ca II,TiO | 8520-8564 | 5.1 | 5.6 | 5.1 | 6.0 | 6.2 | 3.3 | 4.6 | 6.4 | 6.1 |
| 80 | Ca II | 8640-8700 | 4.0 | 4.3 | 4.0 | 5.1 | 4.6 | 3.1 | 4.0 | 5.2 | 5.6 |

**Table 4.** Continuum Points

| $\lambda$(Å) | 3 | 11 | 12 | 23 | 26 | G3 | G2 | G1 | G177 |
|---|---|---|---|---|---|---|---|---|---|
| 4020 | 0.58 | 0.61 | 0.54 | 0.68 | 0.40 | 0.74 | 0.63 | 0.52 | 0.50 |
| 4570 | 0.84 | 0.83 | 0.80 | 0.88 | 0.69 | 0.88 | 0.83 | 0.77 | 0.84 |
| 5340 | 0.94 | 0.93 | 0.91 | 0.95 | 0.89 | 0.97 | 0.95 | 0.93 | 0.97 |
| 6630 | 0.95 | 0.96 | 0.95 | 0.96 | 1.04 | 0.94 | 0.95 | 0.95 | 1.00 |
| 6990 | 0.97 | 0.96 | 0.97 | 0.98 | 1.06 | 0.91 | 0.94 | 0.96 | 0.99 |
| 7520 | 1.04 | 0.97 | 1.05 | 1.05 | 1.16 | 0.89 | 0.93 | 0.96 | 1.04 |
| 8040 | 0.98 | 0.92 | 1.04 | 1.04 | 1.15 | 0.86 | 0.91 | 0.96 | 0.98 |
| 8700 | 0.96 | 0.86 | 1.02 | 1.04 | 1.13 | 0.84 | 0.90 | 0.96 | 0.98 |

As a conclusion, we confirm the HGHH92 claim that the metallicity values derived by Frogel (1984) are largely overestimated, likely from zero-point and/or extinction problems. However, the fact that our metallicity estimates agree moderately well with HGHH92's $C - T_1$-based [Fe/H] values argues against extinction as a significant source of uncertainty, since $C - T_1$ is almost as reddening sensitive as $V - K$.

### 4.2. Reddening

Our method for deriving the metallicity is independent of reddening. In turn, the reddening can be determined by comparison of the observed cluster spectral energy distribution with that of a template spectrum of comparable metallicity. Based on the rankings and absolute metallicity values of Sect.4.1, we adopt as reference templates G1 for #26, an average of G1 and G2 for #3, #12 and #23, and G2 for #11. Assuming a normal reddening law (Seaton 1979), we derive the reddening values E(B–V) shown in Table 6. This reddening corresponds to the sum of a foreground (Galactic) one and an internal (NGC 5128) one. None of our clusters is located in the central main body of NGC 5128 (Figure 1). The largest reddening value is that of the most distant cluster #26. It is noticed by Hesser et al. (1984) that the cluster #11 is seen projected on the dust lane; our result suggests that the cluster is located in front of it.

**Table 6.** Metallicity and Total Reddening Estimates

| Cluster | [Z/Z$_\odot$] | E(B–V) |
|---|---|---|
| 3 | -0.27 | 0.03 |
| 11 | -0.44 | 0.03 |
| 12 | -0.19 | 0.06 |
| 23 | -0.30 | 0.03 |
| 26 | -0.08 | 0.13 |

We point out that three of our clusters have E(B–V)=0.03, lower than the uniform values adopted in the photometric studies for all clusters, E(B–V)=0.11 (Frogel 1984) and E(B–V)=0.10 (HGHH92). This deserves some comment: within uncertainties these *absolute* reddening values are comparable, since a difference of 0.07 in E(B–V) may be caused by a ~6% relative error on fluxes, which is the precision of our flux calibration. However, the relative variations within our cluster sample are real, since they were observed in a single run under the same conditions and the clusters have been calibrated in the same way. Thus, cluster #26 suffers from a significant extinction difference of $\Delta$E(B–V)=0.10 with respect to #3, #11 and #23 (Table 6).

We conclude that the source of reddening for the clusters we have observed is mostly foreground, since the more internal clusters exhibit a lower extinction. The transparency appears to vary significantly for different loci. Thus, the adoption of a uniform reddening in the photometric studies for this quite low Galactic latitude galaxy is a source of uncertainty in the metallicity estimates. For globular clusters located in the main body of NGC 5128, where the dust lane dominates, spectrophotometry would certainly be the most appropriate tool, because metallicity and reddening effects can be straightforwardly separated.

We compare, in Figure 4, the relation between star cluster metallicity and radial distance from the galaxy center, in the Galaxy and in NGC 5128. [Z/Z$_\odot$] values of Galactic globular clusters are taken from Bica & Alloin (1986a), ensuring homogeneity in the metallicity scale; radial distances are taken from Djorgovski (1993) for clusters in our Galaxy, and from Hesser et al. (1984) for the NGC

HGHH92 and references therein). Our metallicities being compatible with HGHH92's results for clusters in common (Sect.4.1), we also include their data. Our cluster sample is close to and/or belongs to the upper metallicity envelope between 5 and 15 kpc. Absolute comparison of the radial distributions (Fig.4a and Fig.4b) indicates that metal-rich clusters in NGC 5128 can be found at larger distances than in our Galaxy. However, scaling the data by the galaxy effective radii, $r_e$=4.9 kpc for NGC 5128 (Dufour et al., 1979), and $r_e$=2.7 kpc for the spheroidal component (bulge) of our Galaxy (de Vaucouleurs and Pence, 1978), the distances up to which metal-rich star clusters are detected look rather uniform (Figs. 4c and 4d): Metal-rich globular clusters. are found in the Galaxy up to ~ 3.7 $r_e$, and up to 4.2 $r_e$ in NGC 5128. For further comparison, metal-rich clusters are detected up to 5 $r_e$ in M31 (Huchra et al., 1991), and up to ~ 4.3 $r_e$ in M81 (Lee and Geisler, 1993). NGC 5128 presents peculiarities, like its warped dust lane and shells, which have been for long interpreted as traces of a former merger, the origin of which is however still to be identified (e.g. Thomson 1992). Zepf and Ashman (1993) suggested that elliptical galaxies, if produced by merger of spiral galaxies, should present more spatially extended star cluster systems, with the metal-rich clusters being more centrally concentrated. From Fig 4., and comparison with M31 and M81, metal-rich clusters in NGC 5128 do not appear more centrally concentrated than the metal-poor ones. We conclude that from the spatial distribution it is not clear whether the NGC 5128 globular cluster system could be explained by a merger of two spirals, although the bimodality present in the metallicity histogram suggests this possibility (Zepf & Ashman, 1993). Spectroscopic observations for a large sample of clusters in NGC 5128 are clearly needed in order to clarify this important issue.

## 5. Concluding remarks

We have obtained spectrophotometric data for five globular clusters in NGC 5128. By comparison with metal-rich globular clusters in the Galaxy and in M31, we show that the metallicities of the NGC 5128 globular clusters are not as extreme as previously estimated from infra-red integrated photometry. The most metal-rich globular cluster #26, in our NGC 5128 sample, is comparable to the nearly solar metallicity globular clusters in our Galaxy, NGC 6553 and NGC 6528, and definitely less strong-lined than G177 in M31. The issue whether the globular cluster system in NGC 5128 was enhanced or not in one (or more) merger event(s) would require spectroscopic observations of a large cluster sample, separating age, metallicity and reddening effects.

*Acknowledgements.* We are grateful to the referee, G. Harris, for her numerous interesting comments. We thank the ESO staff at Garching and La Silla for hospitality and support. E.B.

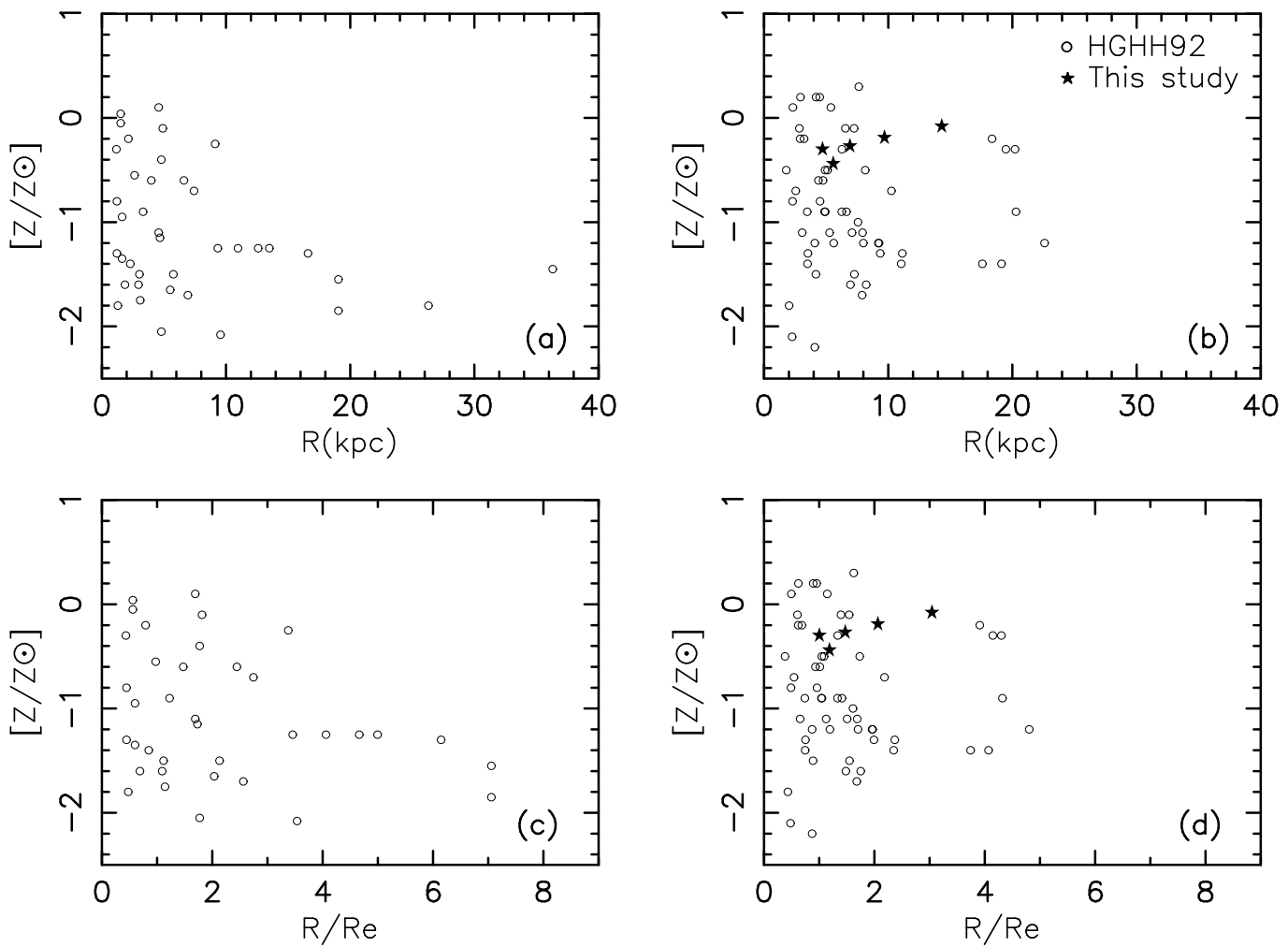

**Fig. 4.** Metallicity distribution of globular clusters (GCs) in our Galaxy (a) and in NGC 5128 (b), as a function of distance from the galaxy center. (c) and (d): the same as respectively (a) and (b), but scaled to effective radius. We assumed d=3.3Mpc for the distance of NGC 5128.

acknowledges the Brazilian sponsoring institutions CNPq and FINEP.